\documentclass[a4paper,12pt]{article}

\usepackage{jcappub} 

\usepackage[T1]{fontenc} %

\usepackage[colorlinks=true,linkcolor=blue,citecolor=blue,urlcolor=blue]{hyperref}

\newcommand{\be}{\begin{equation}}
\newcommand{\ee}{\end{equation}}
\newcommand{\bea}{\begin{eqnarray}}
\newcommand{\eea}{\end{eqnarray}}

\usepackage{psfrag}
\usepackage{latexsym,array,theorem,mathrsfs,subfigure,bm,float}
\usepackage{amsmath}
\usepackage{amsfonts}
\usepackage{amssymb}

\title{Giant wormholes in ghost-free bigravity theory}

\author[a]{Sergey~V.~Sushkov,}
\author[b,a]{Mikhail~S.~Volkov } 

\affiliation[a]{
Department of General Relativity and Gravitation, Institute of Physics,\\
Kazan Federal University, Kremlevskaya street 18, 420008 Kazan, Russia
}

\affiliation[b]{
Laboratoire de Math\'{e}matiques et Physique Th\'{e}orique CNRS-UMR 7350, \\ 
Universit\'{e} de Tours, Parc de Grandmont, 37200 Tours, France}

\emailAdd{\tt serge$\mbox{y}_{-}$sushkov@mail.ru}
\emailAdd{\tt volkov@lmpt.univ-tours.fr}

\abstract{ 
We study Lorentzian wormholes in the ghost-free bigravity theory
described by two metrics, g and f.  
Wormholes can exist if only the null energy condition is violated,
which happens naturally in the bigravity theory 
since the graviton energy-momentum tensors 
do not apriori fulfill any energy conditions. 
As a result, the field 
equations admit solutions describing wormholes whose 
throat size is typically of the order of the 
inverse graviton mass. Hence, they are as large 
as the universe, so that in principle we might all live in a giant wormhole.  
The wormholes can be of two different types that we call W1 and W2. 
The W1 wormholes interpolate between the AdS spaces and have Killing 
horizons shielding the throat.
The Fierz-Pauli graviton mass for these solutions
becomes imaginary in the AdS zone, hence the gravitons behave as tachyons, but since 
the Breitenlohner-Freedman bound is fulfilled, 
there should be no tachyon instability. 
For the W2 wormholes the g-geometry is globally regular and  
in the far field zone it becomes the AdS up to subleading terms,  
its throat can be traversed by timelike geodesics, 
while the f-geometry has a completely different structure
and is not geodesically complete. 
There is no evidence of tachyons for these solutions, although a detailed
stability analysis remains an open issue. It is possible that the solutions may
admit a holographic interpretation. 
 
}

\begin{document}
\maketitle

\flushbottom

\section{Introduction} 
\setcounter{equation}{0}
Lorentzian wormholes are hypothetical field-theory objects describing bridges connecting
different universes or different parts of the same universe. They could supposedly be 
used for momentary displacements over large distances in space. In the simplest case,
a wormhole can be described by a static, spherically symmetric line element 
\be 
ds^2=-Q^2(r)dt^2+dr^2+R^2(r)(d\vartheta^2+\sin^2\vartheta d\varphi^2),
\ee  
where $Q(r)=Q(-r)$ and $R(r)=R(-r)$, both $Q$ and $R$ are positive, and $R$ 
attains a non-zero global 
minimum at $r=0$. If both $Q$ and $R/r$ approach unity at infinity, then 
the metric describes two asymptotically flat regions connected by a throat of 
radius $R(0)$.   Using the Einstein equations $G^\mu_\nu=8\pi G T^\mu_\nu$,
one finds that the energy density $\rho=-T^0_0$ and the 
radial pressure $p=T^r_r$ in the throat at $r=0$ satisfy
\be 
\rho+p=-\frac{R^{\prime\prime}}{4\pi G R}<0,~~~~~
p=-\frac{1}{8\pi G R^2}<0.
\ee
It follows that for the wormhole to be a solution of the Einstein equations, 
the matter should violate the null energy condition ($T_{\mu\nu}v^\mu v^\nu\geq 0$ for any null $v^\mu$). 
This shows that wormholes 
cannot exist in ordinary physical situations where 
the energy conditions are fulfilled. 

However, it was emphasized  
\cite{Morris:1988cz,Morris:1988tu} 
that wormholes could in principle be created by the 
vacuum polarization effects, 
since the vacuum energy can be negative.  
This observation triggered a raise of activity (see \cite{Visser:1995cc} for a review), 
even though 
the wormholes supported by the vacuum effects are typically very small 
\cite{Hochberg:1996ee,Khusnutdinov:2002qb}
and cannot be used for space travels.  
Another possibility to get wormholes is to consider exotic matter types,
as for example phantom fields with a negative kinetic energy 
\cite{Bronnikov:1973fh,Bronnikov:2005gm,Lobo:2005us}.
Otherwise, one can 
search for wormholes in the alternative theories of gravity, as for example in the theories 
with higher derivatives \cite{Hochberg:1990is,Harko:2013yb},
in the Gauss-Bonnet theory \cite{Maeda:2008nz,Kanti:2011jz,Kanti:2011yv,Mehdizadeh:2015jra},
in the brainworld  models 
\cite{Bronnikov:2002rn}, or 
in the Horndeski-type theories \cite{Horndeski:1974wa} with 
non-minimally coupled fields 
\cite{Balakin:2010ar,Sushkov:2011jh,Korolev:2014hwa,Balakin:2014nra}.

A particular case of the Horndeski theory is the Galileon model \cite{Nicolis:2008in},
which can be viewed as a special limit of the ghost-free massive gravity theory
\cite{deRham:2010kj}. This latter theory has recently attracted a lot of attention
(see \cite{Hinterbichler:2011tt,deRham:2014zqa} for a review), 
because it avoids the long standing problem of the ghost \cite{Boulware:1973my}
and could in principle be used to describe the cosmology. In particular, it admits 
self-accelerating cosmological solutions and also black holes 
(see \cite{Volkov:2013roa,Volkov:2014ooa} for a review).   
At the same time, wormholes with massive gravitons have never been
considered. To fill this gap, we shall study below wormholes within the ghost-free
bigravity theory.  

The ghost-free bigravity \cite{Hassan:2011zd} is the extension of the massive gravity theory containing 
two dynamical metrics, $g_{\mu\nu}$ and $f_{\mu\nu}$. They describe two gravitons,
one massive and one massless, and satisfy two coupled sets of Einstein's equations, 
\bea                                  \label{Einstein0}
G^\mu_\nu(g)&=&\kappa_1\, T^{\mu}_{~\nu}(g,f),~~~~~~~~~~~~
G^\mu_\nu(f)=\kappa_2\, {\cal T}^{\mu}_{~\nu}(g,f), 
\eea
where $T^{\mu}_{~\nu}$ and ${\cal T}^{\mu}_{~\nu}$ are the graviton energy-momentum tensors.
What is interesting, these tensors  do not apriori fulfil the 
null energy condition \cite{Baccetti:2012re}, 
which suggests looking for wormhole solutions%
\footnote{Even though the energy 
conditions are not fulfilled, this does  not necessarily 
mean that the energy is negative, and in fact the analysis in the 
massive gravity limit indicates that the energy is positive in the asymptotically flat case 
\cite{Volkov:2014qca,Volkov:2014ida}.}. 

There are two possible ways to interpret the two metrics in 
the theory. One possibility is to view them as describing two geometries on the spacetime manifold. 
Each geometry has its own geodesic structure, and in principle one could introduce two different 
matter types -- a g-matter that follows the g-geodesics and does not directly see the f-metric, 
and an f-matter moving along the f-geodesics. However, it is not always possible to 
put two different geometries on the same 
manifold, and in fact we shall present below solutions for which 
the spacetime manifold is geodesically complete in one geometry but is incomplete
in the other. We shall therefore adopt the viewpoint according to which only the 
$g_{\mu\nu}$  describes the spacetime geometry, 
while the $f_{\mu\nu}$ is a spin-2 tensor 
field whose geometric interpretation is possible but not necessary.

In what follows we shall study the wormhole solutions in the system \eqref{Einstein0}. 
It turns out that such wormholes exist and are gigantic, 
with the throat size of the order of the inverse graviton mass,
which is as large as the universe. Therefore, if the bigravity theory indeed describes the nature,
we might in principle all live in  a giant wormhole.  
We find wormholes of two different types that we call W1 and W2. 
For the W1 solutions both metrics interpolate between the AdS spaces,
and the g-geometry is either globally regular (the W1a subcase)
or it exhibits Killing horizons 
shielding the throat (the W1b subcase). 
The Fierz-Pauli graviton mass  computed in the AdS far field zone
turns out to be imaginary, so that the gravitons behave as tachyons. For the W1a solutions 
the graviton mass violates the Breitenlohner-Freedman (BF) bound, hence these solutions must 
be unstable, but for the W1b solutions the bound is fulfilled, which suggests that the 
tachyon instability is absent. For the W2 solutions the  
 g-metric is globally regular and 
in the far field zone becomes the AdS up to subleading terms,  
while the f-geometry has a completely different structure
and is not geodesically complete. 
There is no evidence of tachyons for these solutions, although a detailed
stability analysis remains an open issue in all cases. It is possible that the solutions may
admit a holographic interpretation.

The rest of the paper is organized as follows. In section \ref{sec2} we introduce the 
ghost-free bigravity theory, whose reduction to the spherically symmetric sector is given 
in section \ref{sec3}. The master field equations and their simplest solutions 
are presented in sections \ref{sec4} and \ref{simp}.  The local solutions 
in the wormhole throat  are obtained in section \ref{sec6},
while section \ref{sec7} presents the global solutions. 
The geometry of the solutions, their global structure,
geodesics, etc, are considered in section \ref{sec7a}.
Other properties of the solutions, in particular their stability,
are briefly discussed in the  final section \ref{sec8}.

\section{The ghost-free bigravity \label{sec2}}
\setcounter{equation}{0}

The theory is  defined on a four-dimensional spacetime manifold endowed with 
two Lorentzian metrics $g_{\mu\nu}$ and $f_{\mu\nu}$ with the signature
$(-,+,+,+)$.   
The action is  \cite{Hassan:2011zd}
\bea                                      \label{1}
&&S
=\frac{M_{\rm Pl}^2}{m^2}\int \, \left(
\frac{1}{2\kappa_1}\,R({g})\sqrt{-{g}}
+\frac{1}{2\kappa_2}\,R({f})\sqrt{-{f}}
-{\cal U}\sqrt{-{g}}\right)\,d^4x
\,,
\eea
where $m$ is the graviton mass and the two gravitational couplings fulfill  
$
\kappa_1+
\kappa_2=1$. 
 The metrics and all coordinates are 
assumed to be dimensionless, with the length scale
being the inverse graviton mass $1/m$.  
 The interaction between the two metrics 
is expressed by  
 a scalar function 
of the tensor 
\be                             \label{gam}
\gamma^\mu_{~\nu}=\sqrt{{{g}}^{\mu\alpha}{{f}}_{\alpha\nu}},
\ee
where ${{g}}^{\mu\nu}$ is the inverse of ${g}_{\mu\nu}$
and the square root
is understood in the matrix sense, i.e. 
\be                     \label{gamgam}
(\gamma^2)^\mu_{~\nu}\equiv \gamma^\mu_{~\alpha}\gamma^\alpha_{~\nu}={g}^{\mu\alpha}
{f}_{\alpha\nu}.
\ee 
If $\lambda_A$ ($A=0,1,2,3$) are the eigenvalues of $\gamma^\mu_{~\nu}$
then the interaction potential is 
\be                             \label{2}
{\cal U}=\sum_{n=0}^4 b_k\,{\cal U}_k(\gamma),
\ee
where $b_k$ are parameters, while  
${\cal U}_k(\gamma)$ are defined 
by the 
relations
\bea                        \label{4}
{\cal U}_0(\gamma)&=&1,  \\
{\cal U}_1(\gamma)&=&
\sum_{A}\lambda_A=[\gamma],\nonumber \\
{\cal U}_2(\gamma)&=&
\sum_{A<B}\lambda_A\lambda_B 
=\frac{1}{2!}([\gamma]^2-[\gamma^2]),\nonumber \\
{\cal U}_3(\gamma)&=&
\sum_{A<B<C}\lambda_A\lambda_B\lambda_C
=
\frac{1}{3!}([\gamma]^3-3[\gamma][\gamma^2]+2[\gamma^3]),\nonumber \\
{\cal U}_4(\gamma)&=&
\lambda_0\lambda_1\lambda_2\lambda_3
=
\frac{1}{4!}([\gamma]^4-6[\gamma]^2[\gamma^2]+8[\gamma][\gamma^3]+3[\gamma^2]^2
-6[\gamma^4])\,. \nonumber 
\eea
Here, using the hat to denote matrices, one has defined 
$[\gamma]={\rm tr}(\hat{\gamma})\equiv \gamma^\mu_{~\mu}$, 
$[\gamma^k]={\rm tr}(\hat{\gamma}^k)\equiv (\gamma^k)^\mu_{~\mu}$. 
The two metrics 
actually enter the action in a completely symmetric way, 
since the action is invariant under 
\be                                  \label{sym}
g_{\mu\nu}\leftrightarrow f_{\mu\nu},~~~~
\kappa_1 \leftrightarrow \kappa_2,~~~~
b_k \leftrightarrow b_{4-k}\,.
\ee 
The parameters $b_k$ can apriori be arbitrary, but
if one requires the flat space to be a solution
of the theory and $m$ to be the Fierz-Pauli
mass of the gravitons in flat space, then 
the five $b_k$ are 
expressed in terms of two arbitrary 
parameters, sometimes called $c_3$ and $c_4$, as 
\be                \label{bbb}
b_0=4c_3+c_4-6,~~
b_1=3-3c_3-c_4,~~
b_2=2c_3+c_4-1,~~
b_3=-(c_3+c_4),~~
b_4=c_4.
\ee
Varying the action (\ref{1}) with respect to 
the two metrics 
gives two sets of the Einstein equations \eqref{Einstein0},
where $T^{\mu}_{~\nu}$ and ${\cal T}^{\mu}_{~\nu}$
are obtained by varying the interaction, 
\bea                        \label{T}
&&
T^{\mu}_{~\nu}=
\,\tau^\mu_{~\nu}-{\cal U}\,\delta^\mu_\nu,~~~~~
{\cal T}^{\mu}_{~\nu}=-\frac{\sqrt{-g}}{\sqrt{-f}}\,\tau^\mu_{~\nu}\,,
\eea
with 
\bea                                \label{tau1}
\tau^\mu_{~\nu}&=&
\{b_1\,{\cal U}_0+b_2\,{\cal U}_1+b_3\,{\cal U}_2
+b_4\,{\cal U}_3\}\gamma^\mu_{~\nu} \nonumber \\
&-&\{b_2\,{\cal U}_0+b_3\,{\cal U}_1+b_4\,{\cal U}_2\}(\gamma^2)^\mu_{~\nu} \nonumber  \\
&+&\{b_3\,{\cal U}_0+b_4\,{\cal U}_1\}(\gamma^3)^\mu_{~\nu} \nonumber \\
&-&b_4\,{\cal U}_0\,(\gamma^4)^\mu_{~\nu}\,.
\eea
The Bianchi identities for the Einstein equations 
imply that 
\be                                   \label{T1} 
\stackrel{(g)}{\nabla}_\rho T^\rho_\lambda=0\,,~~~~~
\stackrel{(f)}{\nabla}_\rho {\cal T}^\rho_\lambda=0\,,~~~~~
\ee
where $\stackrel{(g)}{\nabla}_\rho$ and 
$\stackrel{(f)}{\nabla}_\rho$ are the covariant 
derivatives with respect to $g_{\mu\nu}$ and $f_{\mu\nu}$.

We consider the vacuum theory, but one could also add a matter. 
The equivalence principle and the absence of the ghost require the matter to 
be coupled to one of the two metrics but not 
to both of them at the same time. One could also introduce a g-matter for the g-metric
and an f-matter for the f-metric.  This is important in what follows: test g-particles 
will follow geodesics of the g-metric and will not directly feel the f-metric.

\section{Spherical symmetry \label{sec3}}
\setcounter{equation}{0}
Let us choose both metrics to be spherically symmetric,  
\bea                             \label{ansatz0}
ds_g^2&=&-Q^2dt^2+\frac{dr^2}{\Delta^2}+R^2d\Omega^2\,, \nonumber \\
ds_f^2&=&-q^2 dt^2+\frac{dr^2}{W^2}+U^2d\Omega^2,
\eea
where $Q,\Delta,R,q,W,U$ depend on the radial coordinate $r$ and 
$d\Omega^2=d\vartheta^2+\sin^2\vartheta d\varphi^2$\,. 
In this case the tensor $\gamma^\mu_{~\nu}$ in \eqref{gam} becomes 
\be 
\gamma^\mu_{~\nu}={\rm diag}\left[
\frac{q}{Q},\frac{\Delta}{W},\frac{U}{R},\frac{U}{R}
\right].
\ee
The formulas \eqref{T} then give 
\bea 
T^\mu_{~\nu}&=&{\rm diag}\left[
T^0_0,T^1_1,T^2_2,T^2_2
\right],\nonumber \\
{\cal T}^\mu_{~\nu}&=&{\rm diag}\left[
{\cal T}^0_0,{\cal T}^1_1,{\cal T}^2_2,{\cal T}^2_2
\right],
\eea
where 
\bea
T^0_0&=&-{\cal P}_0-{\cal P}_1\,\frac{\Delta}{W}, \nonumber \\
T^1_1&=&-{\cal P}_0-{\cal P}_1\,\frac{q}{Q},~~\nonumber \\
T^2_2&=&-{\cal D}_0-{\cal D}_1\left(\frac{q}{Q}
+\frac{\Delta}{W}\right)-{\cal D}_2\,
\frac{q\Delta}{QW},\nonumber \\
u^2{\cal T}^0_0&=&-{\cal P}_2-{\cal P}_1\,\frac{W}{\Delta}, \nonumber \\
u^2{\cal T}^1_1&=&-{\cal P}_2-{\cal P}_1\,\frac{Q}{q},~\nonumber \\
u{\cal T}^2_2&=&-{\cal D}_3-{\cal D}_2\left(\frac{Q}{q}
+\frac{W}{\Delta}\right)-{\cal D}_1\,
\frac{QW}{q\Delta}. 
\eea
Here $u={U}/{R}$ and 
\bea 
{\cal P}_m&=&b_m+2b_{m+1}u+b_{m+2}u^2\,, \nonumber \\ 
{\cal D}_m&=&b_m+b_{m+1}u\,~~~~              ~~~(m=0,1,2). 
\label{e5} 
\eea
As one can see, the energy-momentum tensors do not 
apriori 
fulfill any positivity conditions. 
The independent field equations are 
\bea                                  \label{Ein}
G^0_0(g)&=&\kappa_1\, T^{0}_{0}, ~~~~\nonumber \\
G^1_1(g)&=&\kappa_1\, T^{1}_{1}, ~~~~
\nonumber \\
G^0_0(f)&=&\kappa_2\, {\cal T}^{0}_{0},~~~~\nonumber \\
G^1_1(f)&=&\kappa_2\, {\cal T}^{1}_{1},~
\eea
plus the conservation condition 
$
\stackrel{(g)}{\nabla}_\mu T^\mu_\nu=0\,,
$
which has only one non-trivial component, 
\be                                \label{CONS}
\stackrel{(g)}{\nabla}_\mu T^\mu_r=
\left(T^1_1\right)^\prime
+\left.\left.\frac{Q^\prime}{Q}\right(T^1_1-T^0_0\right)
+2\left.\left.\frac{R^\prime}{R}
\right(T^1_1-T^2_2\right)=0.
\ee
The
conservation condition for the second energy-momentum tensor 
also has only one non-trivial component,  
\be                                \label{CONSf}
\stackrel{(f)}{\nabla}_\mu {\cal T}^\mu_r=
\left({\cal T}^1_1\right)^\prime
+\left.\left.\frac{q^\prime}{q}\right({\cal T}^1_1-{\cal T}^0_0\right)
+2\left.\left.\frac{U^\prime}{U}
\right({\cal T}^1_1-{\cal T}^2_2\right)=0,
\ee
but this condition is not independent and actually follows from \eqref{CONS}. 
As a result, there are  5 independent equations \eqref{Ein},\eqref{CONS},
which is enough to determine the 6 field amplitudes $Q,\Delta,R,q,W,U$, 
 because  
 the freedom of 
reparametrizations  of the radial coordinate
$
r\to \tilde{r}(r)
$
allows one to fix one of the amplitudes. 

\section{Field equations \label{sec4}}
\setcounter{equation}{0}

Let us introduce new functions 
\be                                 \label{NY}
N=\Delta R^\prime\,,~~~~Y=WU^\prime\,,
\ee
in terms of which the 
 two metrics read 
\bea                             \label{ansatz000}
ds_g^2&=&-Q^2dt^2+\frac{dR^2}{N^2}+R^2d\Omega^2\,, \nonumber \\
ds_f^2&=&-q^2 dt^2+\frac{dU^2}{Y^2}+U^2d\Omega^2.
\eea
The advantage of this parametrization 
is that 
the second derivatives disappear from the Einstein tensor, and 
the four Einstein equations \eqref{Ein} become 
\bea
N^\prime&=&-\frac{\kappa_1}{2}\frac{R}{NY}\left(R^\prime Y{\cal P}_0
+U^\prime N {\cal P}_1
 \right)+\frac{(1-N^2)R^\prime}{2RN}\,,  \label{e1} \\
Y^\prime&=&-\frac{\kappa_2}{2}\frac{ R^2}{UNY}
\left(R^\prime Y {\cal P}_1+U^\prime N {\cal P}_2
 \right)+\frac{(1-Y^2)U^\prime}{2UY}\,, \label{e2}  \\
Q^\prime&=&-\left(
\kappa_1(Q{\cal P}_0+q{\cal P}_1)+\frac{Q(N^2-1)}{R^2}
\right) \frac{RR^\prime}{2N^2}\,, \label{e3}  \\
q^\prime&=&-\left(
\kappa_2(Q{\cal P}_1+q{\cal P}_2)+\frac{q(Y^2-1)}{R^2}
\right) \frac{R^2U^\prime}{2Y^2U}\,. \label{e4} 
\eea
The conservation condition \eqref{CONS} reads 
\bea                                \label{CONSg}
\stackrel{(g)}{\nabla}_\mu T^\mu_r(g)&=&
\frac{U^\prime}{R}\left(1-\frac{N}{Y}\right)\left(d{\cal P}_0+
\frac{q}{Q}\,d{\cal P}_1\right)
+\left(
\frac{q^\prime}{Q}-\frac{NQ^\prime U^\prime}{YQR^\prime}
\right){\cal P}_1=0,
\eea
and using Eqs.\eqref{e3},\eqref{e4}, this 
reduces to 
\bea                                \label{CONSgg}
R^2Q\stackrel{(g)}{\nabla}_\mu T^\mu_r(g)&= &
\frac{U^\prime}{Y}\,{\bf C}=0\,,
\eea
where
\bea                             \label{C}
{\bf C}&=&\left(
\kappa_2\,\frac{R^4{\cal P}_1^2}{2UY}
-\kappa_1\,\frac{R^3\,{\cal P}_0{\cal P}_1}{2N}
-\frac{(N^2-1)\,R{\cal P}_1}{2N}+(N-Y)Rd{\cal P}_0
\right)Q \nonumber \\
&+&\left(
\kappa_2\,\frac{R^4{\cal P}_1{\cal P}_2}{2UY}
-\kappa_1\,\frac{R^3\,{\cal P}_1^2 }{2N}
+\frac{(Y^2-1)\,R^2{\cal P}_1}{2UY}+(N-Y)Rd{\cal P}_1
\right)q\,,
\eea
with
\be  
d{\cal P}_m=
2\,(b_{m+1}+b_{m+2}u)~~~(m=0,1).          \label{e5a} 
\ee
The conservation condition \eqref{CONSf} becomes 
\be                                \label{CONSff}
-U^2q\stackrel{(f)}{\nabla}_\mu T^\mu_r(f)=
\frac{R^\prime }{N}\,{\bf C}=0\,.
\ee
The two conditions \eqref{CONSgg} and \eqref{CONSff} together require that 
either $U^\prime=R^\prime=0$, in which case 
both metrics are degenerate, or that 
\be                          \label{CC}
{\bf C}=0. 
\ee 
As a result, we obtain the four differential equations \eqref{e1}--\eqref{e4} 
plus the algebraic constraint \eqref{CC}. The  same equations 
can be obtained by inserting the metrics \eqref{ansatz000} directly to the 
action \eqref{1}, which gives 
\bea                                      \label{1a}
&&S
=4\pi \frac{M_{\rm Pl}^2}{m^2}\int L\, dt dr
\,,
\eea
where, dropping the total derivative, 
\bea 
L&=&\frac{1}{\kappa_1}\left(
\frac{(1-N^2)\, R^\prime}{N}-2RN^\prime
\right)Q
+\frac{1}{\kappa_2}\left(
\frac{(1-Y^2)\,U^\prime }{Y}-2UY^\prime
\right)q \nonumber \\
&-&\frac{QR^2 R^\prime }{N}\,{\cal P}_0
-\left(
\frac{QR^2 U^\prime}{Y}+\frac{qR^2R^\prime}{N}
\right){\cal P}_1
-\frac{qR^2 U^\prime }{Y}\,{\cal P}_2\,.
\eea
Varying $L$ with respect to $N,Y,Q,q$ gives 
Eqs.\eqref{e1}--\eqref{e4}, while varying it with respect to $R,U$ 
reproduces conditions  \eqref{CONSgg} and \eqref{CONSff}.  
The  
equations and the Lagrangian $L$ are invariant under the interchange symmetry 
\eqref{sym}, which now reads 
\be                                 \label{change}
\kappa_1\leftrightarrow\kappa_2,~~Q\leftrightarrow q,~~~
N\leftrightarrow Y,~~~R \leftrightarrow U,~~~ b_m\leftrightarrow b_{4-m}\,.
\ee
Equation \eqref{e1}--\eqref{e4} contain $U^\prime$, 
but so far the expression for $U^\prime$ is missing. To obtain it, the 
only way is to differentiate the constraint, 
which gives 
\be 
\frac{\partial \bf C}{\partial N}\,N^\prime+
\frac{\partial \bf C}{\partial Y}\,Y^\prime+
\frac{\partial \bf C}{\partial Q}\,Q^\prime+
\frac{\partial \bf C}{\partial q}\,q^\prime+
\frac{\partial \bf C}{\partial R}\,R^\prime+
\frac{\partial \bf C}{\partial U}\,U^\prime=0.
\ee 
Since the derivatives $N^\prime$, $Y^\prime$, $Q^\prime$, $q^\prime$ expressed by  
Eqs.\eqref{e1}--\eqref{e4} are linear functions of $U^\prime$, this gives a linear 
in $U^\prime$ relation, which can be resolved to yield 
\be                                \label{U1}
U^\prime={\cal D}_U(N,Y,Q,q,R,R^\prime,U).
\ee 
This equation and Eqs.\eqref{e1}--\eqref{e4}   comprise together a closed system
of five differential equations for five variables $N,Y,Q,q,U$. The $R$-amplitude 
is determined by fixing the gauge, for example $R=r$ or $R^\prime=N$. 
One can integrate the  five differential equations by imposing 
the constraint ${\bf C}=0$ only on the initial values, and then it will be fulfilled
everywhere. 

Alternatively, one can integrate only the four equations \eqref{e1}--\eqref{e4}
assuming that $U^\prime$ in their right hand side is given by \eqref{U1}, while 
$U$ 
is obtained by resolving the constraint.  

Yet one more possibility is to use the fact that the constraint 
is linear in $Q,q$. Therefore, it can be resolved with respect to $q$,  
\be                \label{q}
q=\Sigma(N,Y,R,U)\,Q.
\ee
Injecting this to Eqs.\eqref{e1},\eqref{e2},\eqref{U1} gives a closed 
system of three differential equations 
\bea
N^\prime&=&{\cal D}_N(N,Y,U,R,R^\prime),\nonumber \\
Y^\prime&=&{\cal D}_Y(N,Y,U,R,R^\prime), \nonumber \\
U^\prime&=&{\cal D}_U(N,Y,U,R,R^\prime),  \label{eqs}
\eea
and when their solution is known, the amplitude $Q$ 
is obtained from equation \eqref{e3} which assumes the form 
\be                \label{Q1}
Q^\prime=FQ
\ee
with
\be                      \label{F}
F=-\left(
\kappa_1({\cal P}_0+\Sigma{\cal P}_1)+\frac{N^2-1}{R^2}
\right) \frac{RR^\prime}{2N^2}\,.
\ee

\section{Simplest solutions \label{simp}} 
\setcounter{equation}{0}
Some simple solutions of the field equations can be 
obtained analytically \cite{Volkov:2012wp},\cite{Hassan:2012wr}. 
They can be of two different types described below in this section. 
They are not of the wormhole type, but the wormholes constructed 
in the next sections approach these solutions in the far field zone.

\subsection{Proportional backgrounds} 

Let us choose 
the two metrics to be conformally related \cite{Volkov:2012wp},\cite{Hassan:2012wr},
\be                           \label{prop1}
ds_f^2=\lambda^2 ds_g^2\,,
\ee
with a constant $\lambda$. 
This implies that 
\be                           \label{prop2}
q=\lambda Q,~~~U=\lambda R,~~~~~Y=N\,.
\ee 
This also implies that ${\cal P}_m={\cal P}_m(\lambda)$ are constant. 
Imposing the Schwarzschild gauge, 
$
R^\prime=1,
$
the field equations \eqref{e1}--\eqref{e4} and the constraint \eqref{CC} reduce to 
\be                                         \label{conf}
\left(RN^2\right)^\prime=1-\kappa_1({\cal P}_0+\lambda{\cal P}_1)R^2,~~~~~
\left(\frac{Q}{N}\right)^\prime=0\,,
\ee 
and to the condition for $\lambda$, 
\be                                 \label{sgm}
\kappa_1({\cal P}_0+\lambda{\cal P}_1)=
\frac{\kappa_2}{\lambda}({\cal P}_1+\lambda{\cal P}_2)\equiv \Lambda(\lambda). 
\ee 
This is an algebraic equation which can have up to four real roots. 
Choosing a root $\lambda$,  
the solution of \eqref{conf} is 
\be                                       \label{eqprop}
N^2= 1-\frac{2M}{R}-\frac{\Lambda(\lambda)}{3}\,R^2\,,~~~~
Q=const.\times N,
\ee
where $M$ is an integration constant. 
Depending on value of $\Lambda(\lambda)$, this
corresponds either to the Schwarzschild or to Schwarzschild-(anti)-de Sitter geometry. 
If the parameters $b_k$ are chosen according to \eqref{bbb},
then the equation \eqref{sgm} always has a root $\lambda=1$, in which case 
$\Lambda(\lambda)=0$.

\subsection{Deformed AdS}

Let us set in the equations 
$U^\prime=q^\prime=0$ \cite{Volkov:2012wp}. This solves Eqs.\eqref{e4},\eqref{CONSgg}, 
while Eqs.\eqref{e1}--\eqref{e3} reduce (with $R^\prime=1$) to 
\bea 
\left(RN^2 \right)^\prime&=& 1-\kappa_1 R^2{\cal P}_0\,,  \nonumber \label{w1}  \\
\left(
\frac{Q}{N}
\right)^\prime
&=&-\kappa_1 q\,\frac{R{\cal P}_1}{2N^3}\,,  \nonumber \\
Y^\prime&=&-\frac{\kappa_2 R^2}{2UN}\,{\cal P}_1\,,  
\eea 
whose solution is 
\bea                           \label{NQ}
N^2&=&1-\kappa_1b_2U^2-\frac{2M}{R}-\kappa_1b_1UR-\frac{\kappa_1 b_0}{3}\,R^2 \,, \nonumber \\
Q&=&\frac{\kappa_1 q}{2}\,N\int_{R}^\infty \frac{R{\cal P}_1}{N^3}\,dR+AN\, , \nonumber \\
Y&=&-\int_0^R \frac{\kappa_2 R^2}{2UN}\,{\cal P}_1\, dR+Y_0\,,
\eea
where $M,A,Y_0$ are integration constants. 
Interestingly, these expressions can describe a wormhole geometry, because 
if $b_0<0$ then   $N^2\to +\infty$ 
for $R\to\infty$, while the constant $M$ can be chosen such that $N^2$ vanishes at 
$R=h$ and $N^2>0$ for $R>h$.  Introducing the radial coordinate
\bea                           \label{NQ1}
r&=&\int_{h}^R\frac{dR}{N(R)}\,
\eea
and setting in \eqref{NQ} $A=0$,   
the g-metric becomes 
$$
ds_g^2=-Q^2dt^2+dr^2+R^2d\Omega^2\,,  
$$
where $R(r)=h+\alpha r^2+\dots$
with $\alpha>0$ and $Q(r)=Q(0)+{\cal O}(r^2)$. This is the wormhole geometry.
Unfortunately, Eq.\eqref{NQ} does not describe an exact solution,
because   the constraint   ${\bf C}=0$ is not fulfilled and so   
the  conservation condition \eqref{CONSff} for the f-metric is not satisfied. 
However, the leading terms in Eq.\eqref{NQ}  describe 
 the asymptotic  form of a more general 
solution whose amplitudes $U,q$ are not identically constant but 
approach constant values at large $R$. Specifically, expanding the field 
equations at large $R$, one finds the following asymptotic solution,  
 \bea                           \label{NQQQ}
N^2&=&
-\kappa_1\frac{b_0}{3}\,R^2
-\kappa_1 b_1U_\infty R  +{\cal O}(1)\equiv N_\infty^2
+{\cal O}(1)
 \,, \nonumber \\
Y&=&-\frac{\sqrt{3}\kappa_2 b_1}{4U_\infty\sqrt{-\kappa_1 b_0} }\,R^2+{\cal O}(R)
\equiv Y_\infty +{\cal O}(R)
, \nonumber \\
Q&=&\frac{q_\infty }{4U_\infty}\, R+{\cal O}(1)\equiv Q_\infty  
+{\cal O}(1) , \nonumber \\
U&=&U_\infty+{\cal O}\left(\frac{1}{R}\right), \nonumber \\
q&=&q_\infty+{\cal O}\left(\frac{1}{R}\right),
\eea
with constant $U_\infty$, $q_\infty$. 
Comparing $N^2$ with $N_{\rm AdS}^2=1-\Lambda R^2/3$ where $\Lambda=-\kappa_1 b_0<0$, 
one can see that the g-metric is the AdS in the leading order, but the subleading 
terms do not have the AdS structure.  

We shall see below that the wormholes 
approach for $R\to\infty$ either  
the proportional AdS solutions \eqref{prop2},\eqref{sgm},\eqref{eqprop}
or the deformed AdS solutions \eqref{NQQQ}. 
We shall call these wormholes, respectively, type W1 and type W2.

\section{Wormholes -- local behavior  \label{sec6} } 
\setcounter{equation}{0}

Since we are unable to obtain the wormhole 
solutions analytically, we resort to the numerical analysis. 
As a first step, we impose the reflection symmetry. 
Let us return for a moment to  the 
parametrization \eqref{ansatz0} 
and require the two metrics 
to be symmetric under $r\to -r$,   
\bea 
Q(r)&=&Q(-r),~~~\Delta(r)=\Delta(-r),~~~R(r)=R(-r), \nonumber \\
q(r)&=&q(-r),~~~W(r)=W(-r),~~~U(r)=U(-r). 
\eea 
Passing then to the parametrization \eqref{NY}, it follows that 
the functions $N,Y$ defined by 
\eqref{NY} should be antisymmetric, 
\be                              \label{YYY}
N(r)=-N(-r),~~~Y(r)=-Y(-r).
\ee
This suggests a local power-series solution around $r=0$, 
\bea                  \label{R}
N&=&N_1r+N_3 r^3+\ldots\,~~~
Q=Q_0+Q_2 r^2+\ldots\,~~~~
R=h+R_2 r^2+\ldots\, \nonumber \\ 
Y&=&Y_1r+Y_3 r^3+\ldots\,~~~~~
q=q_0+q_2 r^2+\ldots\,~~~~~~
U=\sigma h+U_2 r^2+\ldots\, .
\eea
Here $h=R(0)$ is the radius of the wormhole throat 
measured by the first metric, and $\sigma=U(0)/R(0)$ is the 
ratio of the throat  radii measured by the two metrics.  

From now on we shall adopt the gauge condition 
\be 
N=R^\prime\,,
\ee 
which implies that 
\bea                             \label{ansatz0000}
ds_g^2&=&-Q^2dt^2+dr^2+R^2d\Omega^2\,, 
\eea
so that $r$ is the proper distance 
measured by the g-metric. 
The next step is to impose the 
field equations to determine the coefficients in \eqref{R}. 
To begin with, one notices that when inserting \eqref{R} to 
Eqs.\eqref{e1}--\eqref{e4}, \eqref{CC} and expanding the result over $r$,
the leading terms are given by Eqs.\eqref{e3},\eqref{e4} which contain 
a pole in the right hand side due to the terms $R^\prime/N^2\sim 1/r$
and $U^\prime/Y^2\sim 1/r$. For the equations to be fulfilled in the leading order,
the coefficient in front of the pole should vanish, which imposes the conditions 
\bea                           \label{linear1}
\left(\kappa_1{\rm P}_0-\frac{1}{h^2}\right)Q_0
+\kappa_1{\rm P}_1\,q_0&=&0,\nonumber \\
\left(\kappa_2{\rm P}_2-\frac{1}{h^2}\right)q_0
+\kappa_2{\rm P}_1\,Q_0&=&0,
\eea 
with 
$
{\rm P}_m={\cal P}_m(\sigma)=
b_m+2b_{m+1}\sigma
+b_{m+2}\,\sigma^2\,.
$
These two linear equations 
will have a non-trivial solution if only their determinant  
vanishes. Therefore, one requires that 
\be                          \label{D1} 
\left(\kappa_1 h^2{\rm P}_0-1\right)
\left(\kappa_2 h^2{\rm P}_2
-1\right)
-\kappa_1\kappa_2h^4{\rm P}_1^2=0. 
\ee 
If this condition is fulfilled then the solution of \eqref{linear1} is 
\be 
q_0=\alpha Q_0
\ee
with 
\be 
\alpha=\frac{1-\kappa_1 h^2{\rm P}_0}{\kappa_1 h^2 {\rm P}_1}=
\frac{\kappa_2 h^2 {\rm P}_1}{1-\kappa_2 h^2{\rm P}_2}. 
\ee
The value of $Q_0$ is irrelevant, as it can be changed by rescaling the time, 
hence we set $Q_0=1$. 
Eq.\eqref{D1} plays the key role in our analysis and provides the necessary 
condition for the wormholes to exist. For a given $h$, this is a fourth
order algebraic equation for $\sigma$. If we assume for a moment 
that the parameters $b_k$ are given by \eqref{bbb} with $c_3=c_4=0$, 
then the equation becomes quadratic and gives 
\be 
\sigma=\frac{3\kappa_1 h(\kappa_2 h^2-1)\pm \sqrt{\kappa_1(\kappa_2 h^2+1)(3h^2-1)}}
{\kappa_1 h(3\kappa_2 h^2-1)}.
\ee 
This expression, assuming both $\kappa_1$ and $\kappa_2$ to be positive, will be 
real-valued if the square root is real, which requires that $h\geq 1/\sqrt{3}$.   
Since $h$ is measured in unites of $1/m$, which is 
of the order of the Hubble radius, it follows that the wormholes 
are gigantic, as large as the universe. 

Let us return to the expansion of the equations over $r$. Having removed the $r^{-1}$ terms 
in Eqs.\eqref{e3},\eqref{e4}, the 
next-to-leading order terms are provided by Eqs.\eqref{e1},\eqref{e2},
which reduce in the $r^0$ order to 
\bea                     \label{NY1}
N_1&=&-\frac{\kappa_1}{2}\,h\left({\rm P}_0
+\frac{2U_2}{Y_1}\,{\rm P}_1\right)
+\frac{1}{2h}\,,\nonumber \\
Y_1&=&-\frac{\kappa_2}{2}\,\frac{h}{\sigma}
\left({\rm P}_1
+\frac{2U_2}{Y_1}\,{\rm P}_2\right)+\frac{U_2}{\sigma h Y_1}\,.
\eea
These relations can be used to express $R_2=N_1/2$ and $U_2$ in terms of $Y_1$, 
while Eqs.\eqref{e3},\eqref{e4} considered in the $r^1$ order provide 
similar expressions for $Q_2$ and $q_2$. 
Altogether this gives  
\bea                  \label{par0}
N_1&=&2R_2=-\frac{\kappa_1\sigma\alpha}{\kappa_2}\,Y_1,\nonumber \\
U_2&=&\frac{\alpha}{2}\left(
1+\frac{2\sigma Y_1}{\kappa_2 h {\rm P}_1}
\right)Y_1, \nonumber \\
Q_2&=&-\left(\frac{\kappa_1}{4}\left(\frac{2U_2}{Y_1} 
-\sigma\right)(d{\rm P}_0+\alpha d{\rm P}_1)+\frac{R_2}{h}
+\frac{1}{2h^2}\right)Q_0,\nonumber \\
q_2&=&\frac{U_2}{Y_1}\left(
2Q_2+\frac{2R_2-Y_1}{h{\rm P}_1}(d{\rm P}_0+\alpha d{\rm P}_1)Q_0
\right),
\eea
whereas the value of $Y_1$ is fixed by the constraint \eqref{CC}, 
\be                       \label{par1}
Y_1=
\frac{\kappa_2 h^2{\rm P}_1(\kappa_2+\kappa_1\sigma^2)
(
d{\rm P}_0+2\alpha d{\rm P}_1+\alpha^2 d{\rm P}_2
)
-2\sigma{\rm P}_1(\kappa_2+\kappa_1\alpha^2)
}
{
2\sigma h\,[
(\kappa_2+2\kappa_1\alpha\sigma)d{\rm P}_0+2\kappa_1\sigma\alpha^2d{\rm P}_1
-\kappa_2\alpha^2d{\rm P}_2]-2\alpha h{\rm P}_1(\kappa_2+\kappa_1\sigma^2)
}.
\ee
Continuing this process would allow one to recurrently determine all higher order  
coefficients in the expansions \eqref{R}. For example, in the next two orders 
Eqs.\eqref{e1}--\eqref{e4} determine $R_4,Q_4,Q_4,q_4$ in terms of $Y_3$, while the 
latter is determined by the constraint. For given values of the couplings 
$\kappa_1$, $\kappa_2$ and $b_k$, the only free parameter in the expansions is the 
wormhole radius $h$, all other coefficients being fixed by the equations.  
Therefore, the wormholes are characterized by only one continuous parameter, their size $h$. 
However, since the algebraic equation \eqref{D1} can have several roots, there could 
be several different wormholes with the same radius $h$. 

It is worth noting that the expressions in \eqref{par0},\eqref{par1} still exhibit 
the interchange symmetry \eqref{change}, even though this is not completely obvious now, 
when the gauge is fixed.  
Indeed, our gauge choice is $g_{rr}=1$, whereas directly interchanging the metrics 
would give
the solutions in the different gauge, where $f_{rr}=1$. Let us introduce 
the radial coordinate 
\be
z=\int_0^r\frac{U^\prime}{Y}\,dr,
\ee
so that 
the derivative of a function $f$ with respect to $z$ at $r=z=0$ is
$$
f^\prime_z=\frac{Y_1}{2U_2}f^\prime_r \,.
$$
If one interchanges the two metrics, one will have $f^\prime_r\leftrightarrow f^\prime_z$. 
Since $N\leftrightarrow Y$, $R\leftrightarrow U$ and $Q\leftrightarrow q$,  
it follows that the coefficients in \eqref{par0},\eqref{par1} should fulfill the 
relations  
\bea
\frac{Y_1}{2U_2}\,N_1 &=&\underline{Y_1},~~~~~\frac{Y_1}{2U_2}\,Y_1 =\underline{N_1},\nonumber \\
\frac{Y^2_1}{4U^2_2}\,R_2 &=&\underline{U_2},~~~~~
\frac{Y^2_1}{4U^2_2}\,U_2 =\underline{R_2},                   \label{under}
\eea
and similarly for $Q_2$ and $q_2$. Here the underlined expressions
should be evaluated for the interchanged parameter values: $\kappa_1\leftrightarrow \kappa_2$
and $b_k\leftrightarrow b_{4-k}$. A straightforward verification shows that the 
expressions \eqref{par0},\eqref{par1} indeed fulfill
the relations  \eqref{under}. 

\section{Wormholes -- global solutions \label{sec7} } 
\setcounter{equation}{0}

Skipping the important issue of convergence of the power series in \eqref{R},
the above results indicate that the wormhole solutions exist at least locally, 
in the throat. The next step is to construct them globally. 
To this end, 
we extend the local solution \eqref{R} towards large values 
of $r$ numerically, using the 
standard integration  procedure described in \cite{Press:2007:NRE:1403886}. Our results are 
as follows. 

\begin{figure}[h]
\hbox to \linewidth{ \hss

	\psfrag{x}{$\ln(1+r)$}
	\psfrag{R}{$\ln(\ln(R))$}
	\psfrag{Y}{$Y/N$}
	\psfrag{N}{$N/N_0$}	

	\psfrag{Q}{$Q/R$}
	\psfrag{q}{$q/R$}
	\psfrag{U}{$U/R$}	

	\resizebox{8cm}{6cm}
	{\includegraphics{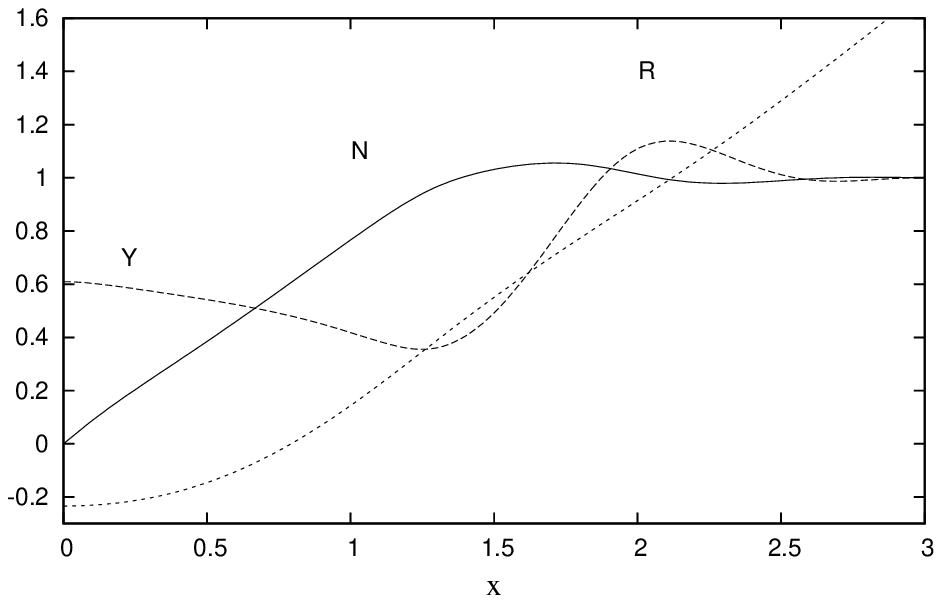}}
\hspace{1mm}
	\resizebox{8cm}{6cm}{\includegraphics{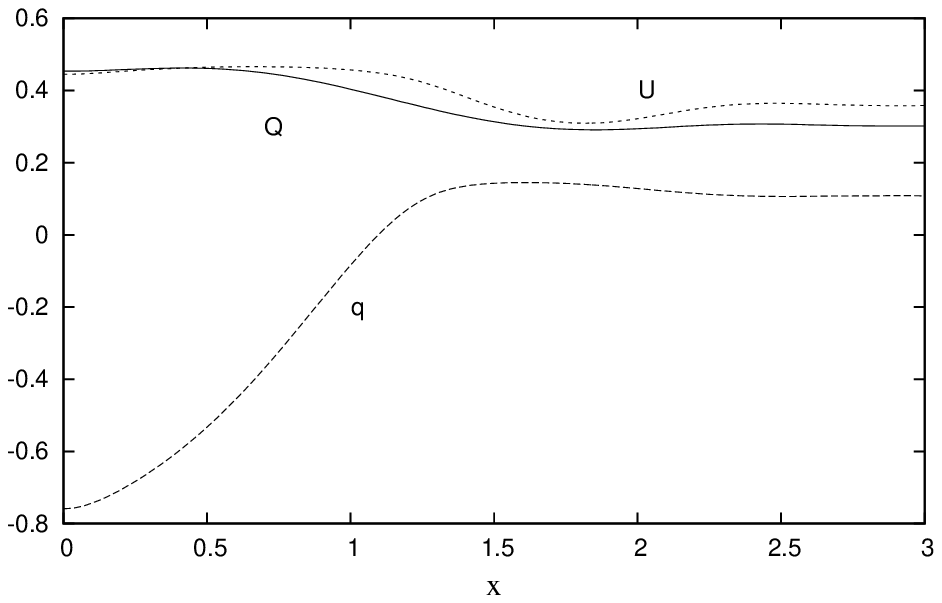}}
	
\hspace{1mm}
\hss}

\caption{{\protect\small 
Profiles of the type W1a wormhole solution obtained for the 
parameter values \eqref{par}. The amplitude $Q$ is everywhere positive 
but $q$ changes sign. For large $r$ the solution approaches 
the proportional AdS background. 
 }}%
 \label{Fig1}
\end{figure}

Choosing some values for 
the parameters $\kappa_1$, $\kappa_2$ and $b_k$, it turns out that the local solution 
\eqref{R}
extends to the whole interval $r\in [0,\infty)$ only for narrow sets of values of 
$h$. These latter are selected by the condition that 
$\sigma$ determined by \eqref{D1} is real. In addition, 
the second derivative $R^{\prime\prime}(0)=2R_2$ should be positive, since otherwise 
$R(r)$ vanishes at a finite value of $r$.  Finally, even if these two conditions are 
fulfilled, the solution may exhibit a singularity at a finite proper distance 
away from the throat at a point where the derivatives $U^\prime$ or $Y^\prime$ diverge.
However, all these problems 
 can be avoided (for some parameter values) by adjusting the throat radius  $h$. 

For properly chosen values of $h$ the solution extends up to large $r$ and 
approaches for $r\to\infty$ either the proportional AdS background or the deformed 
AdS background described in Section \ref{simp}. According to their asymptotic 
behavior, we shall call these wormholes, respectively, either type W1 or type W2.

\subsection{Type W1 wormholes}

For these solutions the two metrics in the far field zone become
proportional to each other and approach 
the proportional AdS background described in Section \ref{simp}. 
However, before reaching  this asymptotic, 
either $Q$ or $q$ or both change sign.
If only one of these amplitudes vanishes, 
then we say that the solution is of type W1a. 
If both $Q$ and $q$ flip sign, then the solution is called type W1b.

\begin{figure}[h]
\hbox to \linewidth{ \hss

	\psfrag{x}{$\ln(1+r)$}
	\psfrag{R}{$\ln(\ln(R))$}
	\psfrag{Y}{$Y/N$}
	\psfrag{N}{$N/N_0$}	

	\psfrag{Q}{$Q/R$}
	\psfrag{q}{$q/R$}
	\psfrag{U}{$U/R$}	

	\resizebox{8cm}{6cm}
	{\includegraphics{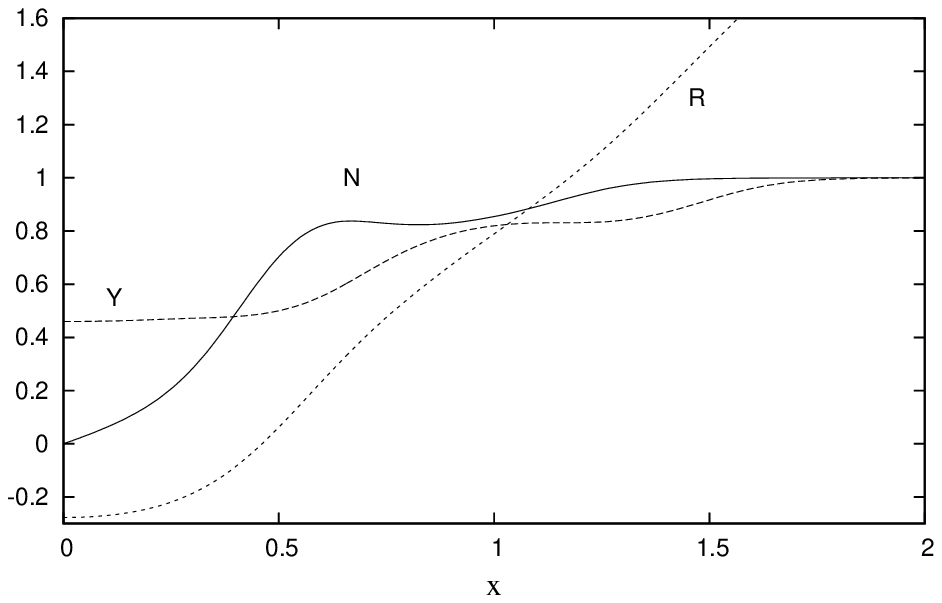}}
\hspace{1mm}
	\resizebox{8cm}{6cm}{\includegraphics{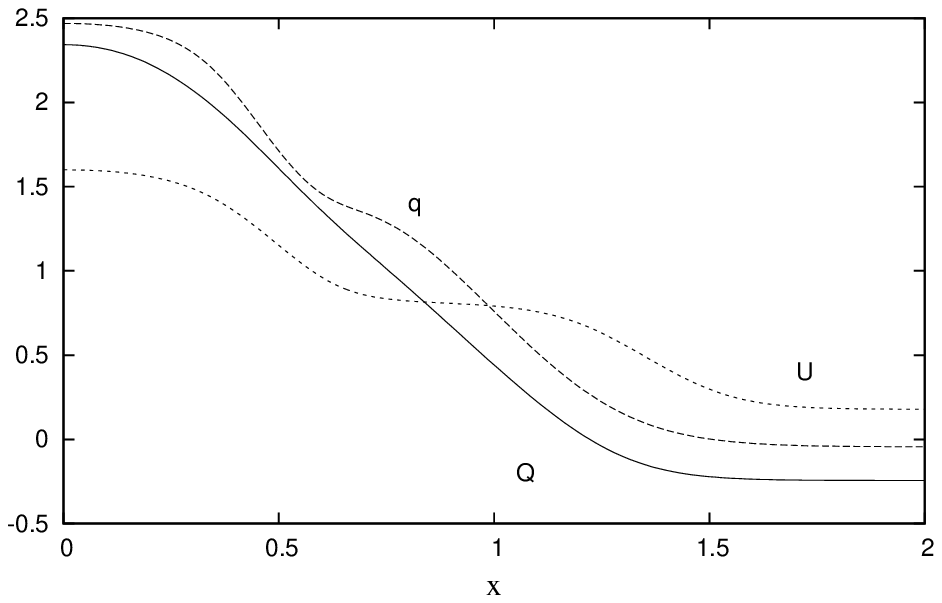}}
	
\hspace{1mm}
\hss}
\caption{{\protect\small 
Profiles of the W1b wormhole solution obtained for the parameter 
values \eqref{par1a}. 
For large $r$ the solution approaches 
the proportional AdS background. 
 Both $Q$ and $q$ change sign. 
 }}%
 \label{Fig1a}
\end{figure}

In Fig.\ref{Fig1}  we present an example of the W1a solution obtained by choosing 
\be                                        \label{par}
\kappa_1=0.688,~~~\kappa_2=0.312,~~~c_3=3,~c_4=-6,~~~h=2.20,~~~~\sigma=0.444,
\ee 
and with $b_k=b_k(c_3,c_4)$ given by \eqref{bbb}. 
The g-metric shows the wormhole throat and is globally regular, 
for large $r$ the whole solution approaching the proportional AdS
background with $f_{\mu\nu}=\lambda^2 g_{\mu\nu}$. To see this, we notice that
assuming the values \eqref{par}, Eq.\eqref{sgm} gives three possible options for the 
proportionality parameter $\lambda$, 
\be                     \label{lam1}
\lambda_1=1,~~~~\lambda_2=-1.264,~~~\lambda_3=0.358,
\ee
with the corresponding values of the cosmological constant   
\be                     \label{lam2}
\Lambda(\lambda_1)=0,~~~~~~\Lambda(\lambda_2)=-7.474,~~~~~\Lambda(\lambda_3)=-0.170.
\ee
The numerical solution chooses the last of these three options as the asymptotic. 
Indeed, the ratio $U/R$ shown in Fig.\ref{Fig1} approaches at large $r$ precisely 
the value $\lambda_3$, as does the ratio $q/Q$, 
while the ratio $Y/N$  (with $N=R^\prime$) approaches the unit value. 
All this agrees with Eq.\eqref{prop2}. Next, according to \eqref{eqprop}, the amplitude 
$N^2$ should approach the AdS value $N_0^2=1-\Lambda(\lambda_3)R^2/3$,
and indeed the ratio $N/N_0$ approaches unity, 
as seen in Fig.\ref{Fig1}. 
Finally, the ratios $Q/R$ and $q/R$ 
should approach constant values, which is indeed the case.

The amplitude $Q$ is everywhere positive, but $q$ changes sign at 
some point, so that 
 the metric coefficient $f_{00}=-q^2$ develops a double zero. 
This corresponds to a Killing horizon of the f-geometry and,
as we shall see below, the curvature diverges at the horizon.
At the same time, nothing special happens to the g-metric at the point 
where $q$ vanishes. The g-geometry is everywhere regular and interpolates between two 
AdS asymptotics  as $r$ varies from $-\infty$ to $+\infty$, passing at $r=0$ through the 
wormhole throat  of size $h$. The test particles of a g-matter coupled to the g-metric 
will therefore see a regular wormhole. 

It is possible that for some special parameter values there could be solutions 
for which both $Q$ and $q$ are sign definite, but  we could not find them. 
On the contrary, we find solutions 
for which both $Q$ and $q$ change sign,
so that both g and f geometries exhibit Killing horizons. The g-horizons and the f-horizons 
are generically located at different points, because they are 
 singular, for if they were regular they would coincide 
to each other \cite{Deffayet:2011rh}. 
An example of the W1b solution obtained for the parameter values 
\be                                        \label{par1a}
\kappa_1=4.446,~~~\kappa_2=-3.446,~~~c_3=1,~c_4=0,~~~h=0.426,~~~~\sigma=1.6
\ee   
is shown in Fig.\ref{Fig1a}.
Notice that $\kappa_2<0$ in this case. 
Since their both metrics show singular horizons,
one could think that the W1b 
solutions are less interesting as compared to the W1a ones.  
However, we shall see below that the W1a 
solutions are prone to the tachyon instability, whereas the W1b solutions seem to be 
free of this problem.

\subsection{Type W2 wormholes}

\begin{figure}[h]
\hbox to \linewidth{ \hss

	\psfrag{x}{$\ln(1+r)$}

	\psfrag{R}{$\ln(\ln(R))$}
	\psfrag{Y}{$Y/Y_\infty$}
	\psfrag{N}{$N/N_\infty$}	

	\psfrag{Q}{$Q/Q_\infty$}
	\psfrag{q}{$\ln(q)$}
	\psfrag{U}{$\ln(U)$}

	\resizebox{8cm}{6cm}
	{\includegraphics{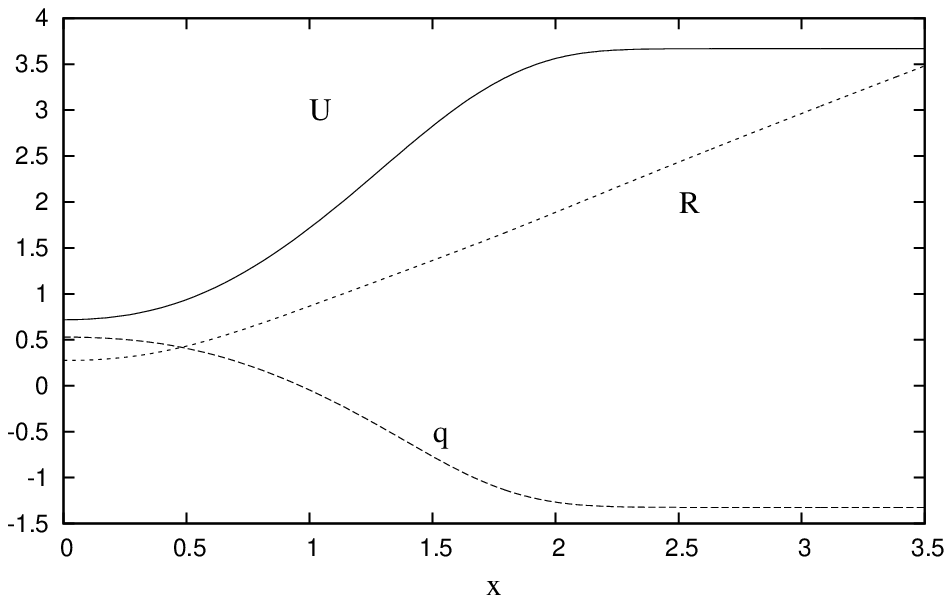}}
\hspace{1mm}
	\resizebox{8cm}{6cm}{\includegraphics{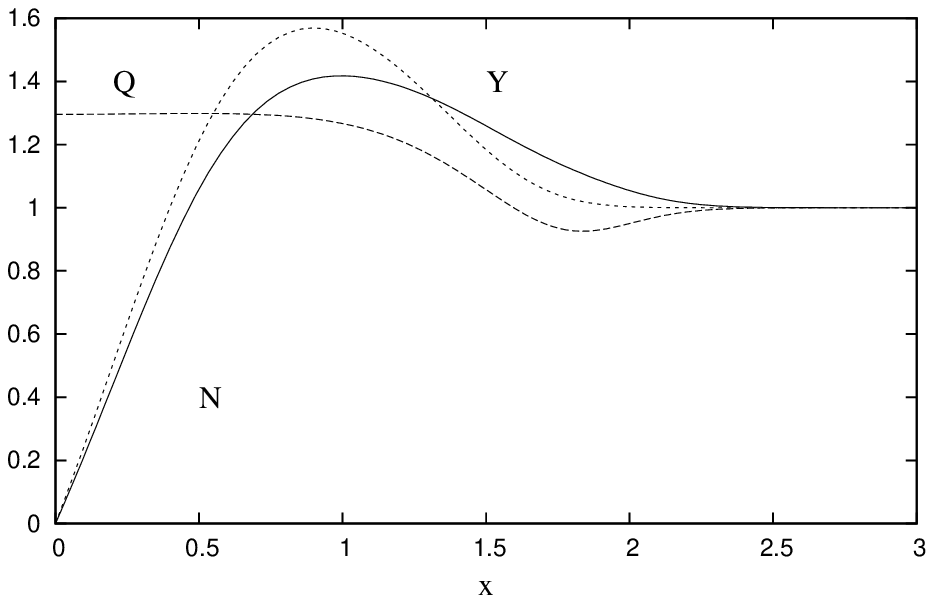}}
	
\hspace{1mm}
\hss}
\caption{{\protect\small 
Profiles of the W2 wormhole solution obtained for the parameter value \eqref{par1b}, with 
$N_\infty$, $Y_\infty$, $Q_\infty$ defined by \eqref{NQQQ}. 
 }}%
 \label{Fig2}
\end{figure}

For these solutions both metrics are globally regular and 
the g-geometry describes a wormhole, but the f-geometry 
is completely different. 
An example of such a solution is shown in
Fig.\ref{Fig2} for the parameter values 
\be                                        \label{par1b}
\kappa_1=0.574,~~~\kappa_2=0.425,~~~c_3=0.1,~c_4=0.3,~~~h=3.731,~~~~\sigma=0.55.
\ee 
For these solutions the amplitudes $U$ and $q$ approach asymptotically 
constant values $U_\infty$ and $q_\infty$. The asymptotic behavior  
of the solutions is described by Eq.\eqref{NQQQ}, which is seen from the 
fact that the ratios $N/N_\infty$, $Y/Y_\infty$, and $Q/Q_\infty$ 
approach unity, with $N_\infty$, $Y_\infty$, $Q_\infty$ defined in \eqref{NQQQ}. 
After a time reparametrization with a constant scale factor, 
the g-geometry in the far field zone is described by  
\be 
ds_g^2=-Q^2 d{t}^2+\frac{dR^2}{N^2}+R^2 d\Omega^2\,,
\ee
where in the leading ${\cal O}(R^2)$ order the $Q,N$ amplitudes coincide with each other and 
with the corresponding amplitude 
for the AdS geometry for the negative cosmological constant $\Lambda=\kappa_1 b_0$,
\be 
Q^2=-\frac{\Lambda}{3}\,R^2+{\cal O}(R),~~~~~N^2=-\frac{\Lambda}{3}\,R^2+{\cal O}(R).
\ee
However, already in the first ${\cal O}(R)$ subleading 
order the amplitudes $Q$ and $N$ are no longer the same 
and deviate from the AdS value. The f-geometry in the asymptotic region 
is expressed by Eq.\eqref{fff} below and
corresponds to a direct product $M^{(1,1)}\times S^2$.

We did not find other global solutions than those of the described above types 
W1 and W2. 
For generic values 
of the parameters $\kappa_1$, $\kappa_2$, $c_3$, $c_4$, $h$ we either do not find any 
solutions at all or obtain singular solutions. If the parameters are properly chosen 
and the solutions exist and 
extend to large values of $r$, then they are always found to be either W1 or W2. 
In addition, due to the symmetry \eqref{change}, there are also solutions 
for which the two metrics are interchanged. A systematic study of the topography 
of the parameter space to identify all parameter values for which the solutions exist 
is a time consuming task that we leave for a future project.

\section{Geometry of the solutions \label{sec7a}}
\setcounter{equation}{0}

Let us consider the geometry of the solutions and its global structure, in order 
to see if the wormholes are traversable or not.

\subsection{Type W1a wormholes}
\begin{figure}[h]
\hbox to \linewidth{ \hss

	\psfrag{x}{$\rho=0$}
	\psfrag{xxx}{$\rho$}
		\psfrag{x1}{$\rho=\rho_\infty$}
	     \psfrag{x2}{$\rho=-\rho_\infty$}	

	\psfrag{light}{\large{light}}
	\psfrag{time}{\large{timelike geodesics}}
	\psfrag{f}{\Large{f-geometry}}
	\psfrag{g}{}

	\psfrag{t}{$t$}
	\psfrag{r}{$\rho$}
	\psfrag{I}{\Large{$J_{+}$}}
	\psfrag{II}{\Large{$J_{-}$}}

	\psfrag{Y}{$Y/Y_\infty$}
	\psfrag{N}{$N/N_\infty$}	

	\psfrag{Q}{$Q/Q_\infty$}
	\psfrag{q}{$\ln(q)$}
	\psfrag{U}{$\ln(U)$}	

	\psfrag{QQ}{$Q^2$}

	\resizebox{4cm}{6cm}
	{\includegraphics{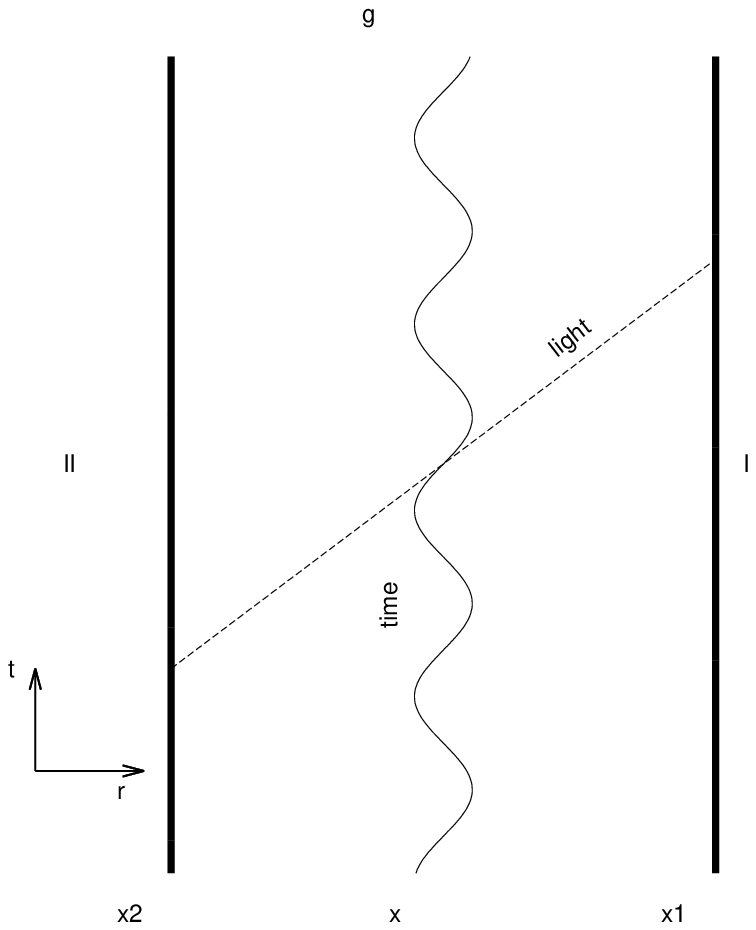}}
\hspace{20mm}
	\resizebox{6cm}{5cm}{\includegraphics{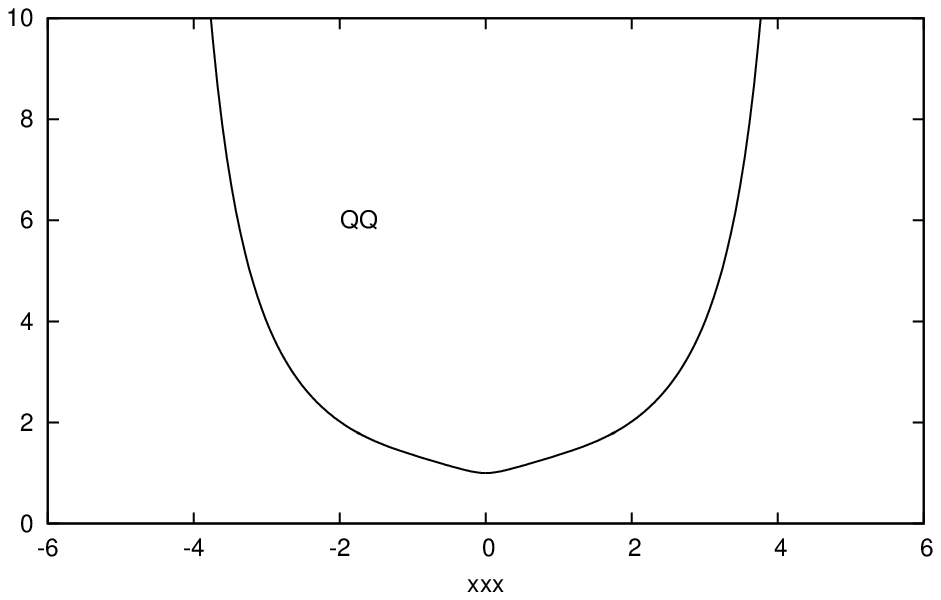}}
	
\hspace{1mm}
\hss}
\caption{{\protect\small 
Conformal structure of the g-geometry for the W1a solution (left) and 
the effective potential $Q^2$    
in the geodesic equation \eqref{geodesic} (right).  
 }}%
 \label{Fig3}
\end{figure}

Let us first consider solutions of the type shown in Fig.\ref{Fig1}. 
One can represent the 2D part of the g-metric as
\be                                    \label{conf1}
ds_g^2=
-Q^2dt^2+dr^2=
Q^2\left(-dt^2+d\rho^2\right)\equiv Q^2 d\bar{s}^2\,.
\ee
Here $r\in(-\infty,+\infty)$ is the proper distance, while 
the conformal radial coordinate 
\be                          \label{rho}
\rho=\int_0^r \frac{dr}{Q}
\ee
changes within a finite interval, $\rho\in(-\rho_\infty,\rho_\infty)$. 
The lightlike geodesics are the same in the $ds_g^2$ and $d\bar{s}^2$ 
geometries. Using the Hamilton-Jacobi equation, 
the radial timelike geodesics followed by particles of mass $\mu$ are described by 
\be                            \label{geodesic}
\left(\frac{d\rho}{dt}\right)^2+\frac{\mu^2}{{\cal E}^2}\,Q^2=1, 
\ee
where ${\cal E}$ is the particle energy. 
Dropping the conformal factor $Q^2$ in \eqref{conf1} 
leads to the conformal diagram of the g-geometry
in the $t,\rho$ coordinates  shown in Fig.\ref{Fig3}.   
The central part of the diagram, $\rho=0,$ is the position
of the throat around which  timelike geodesics  oscillate.  
As is seen in Fig.\ref{Fig3}, the effective potential $Q^2$ has a minimum 
at the throat position, so that the throat attracts the particles. 
The maximal and minimal 
values of the radial coordinate, $\pm\rho_\infty$, correspond to the position of the conformal
timelike boundary, which is at an infinite 
proper distance away from the throat. 
Timelike geodesics are trapped by the confining potential and cannot reach 
the boundary, while  
null geodesics ($\mu=0$) reach it in a finite coordinate time $t$ but 
in an infinite affine time.  
As a result, the g-geometry is geodesically complete. 
All this is very similar to the properties of the  AdS geometry, 
apart from the fact that the boundary 
consists now of two components, $J_{+}$ and   $J_{-}$.  
The conclusion is that timelike geodesics traverse the wormhole 
and oscillate around the throat.

\subsection{Type W1b wormholes}

Let us consider the g-geometry shown in Fig.\ref{Fig1a}.  
The specialty now is that the amplitude $Q$ vanishes at $r=r_\pm$,
where $r_-=-r_+$, 
so that $Q$ is positive for $r\in(r_-,r_+)$ and is negative otherwise. 
The 2D part of the metric can be expressed as
\be
ds_g^2=Q^2(-dt^2+d\rho^2)\equiv -d\tau^2,
\ee 
where the $\rho$-coordinate is defined by \eqref{rho} for  $r\in(r_-,r_+)$,
otherwise one has 
\be                          \label{rho1}
\rho=-\int^{\infty}_r \frac{dr}{Q}~~~~\mbox{if}~~r>r_+~~~~~\mbox{and}~~~~~
\rho=\int_{-\infty}^r \frac{dr}{Q}~~~~\mbox{if}~~r<r_-.
\ee
This determines three coordinate regions:
\bea                             \label{regions} 
A&:&~~~~r\in(r_-,r_+),~~~~~\rho\in (-\infty,\infty), \nonumber \\
B_{+}&:&~~~~r\in (r_+,\infty), ~~~~~~~\rho\in (0,\infty),\nonumber \\
B_{-}&:&~~~~r\in (-\infty,r_-), ~~~~~\rho\in (-\infty,0),
\eea
and in the last two regions $\rho$ changes in the opposite directions, 
so that $\rho=0$ corresponds to $r=\pm\infty$. 
\begin{figure}[h]
\hbox to \linewidth{ \hss
	\psfrag{qq}{\large $q^2$}
	\psfrag{xxx}{$\rho$}

	\psfrag{r}{$\rho=0~~~~~~~r=\infty$}
	\psfrag{time}{\large timelike geodesics}

	\psfrag{I3}{\large ${\cal H}^{-}$}
	\psfrag{I4}{\large ${\cal H}^{+}$}
	\psfrag{I}{\large ${\cal J}$}

	\resizebox{12cm}{5cm}
	{\includegraphics{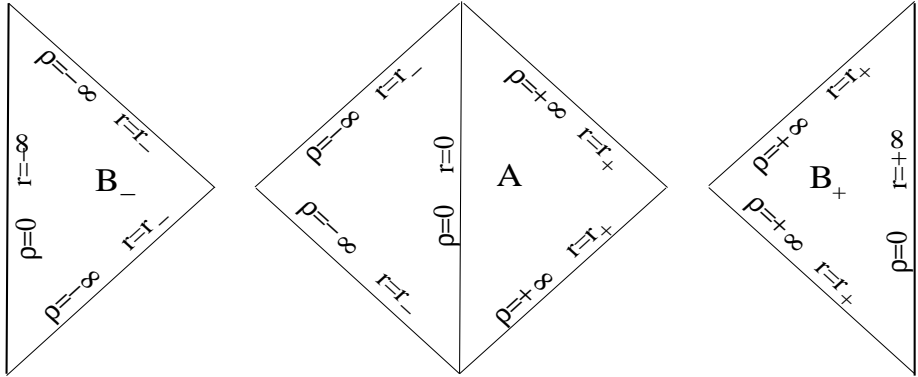}}
	
\hspace{1mm}
\hss}
\caption{{\protect\small 
Conformal structure of the spacetime regions defined by Eq.\eqref{regions}. 
 }}%
 \label{Fig4}
\end{figure}
In the region $A$ one has $\rho\pm t\equiv \tan(u_\pm)\in(-\infty,\infty)$,
so that 
\be                        \label{tau}
ds_g^2=Q^2(-dt^2+d\rho^2)=\frac{Q^2}{\cos^2(u_+)\cos^2(u_-)}\,du_+\,du_-\,,
\ee 
where $u_\pm\in(-\pi/2,\pi/2)$. 
Therefore, the $A$ region is conformally equivalent to the 
diamond in the $(u_+,u_-)$ plane shown 
in Fig.\ref{Fig4}. The vertical symmetry axis of the diamond 
corresponds to the throat position, $\rho=r=0$. In the $B_\pm$ regions
one has either $\rho>0$ or $\rho<0$, hence conformal images of these regions 
can be obtained by cutting the diamond and keeping either only its right or 
only its left triangular part,
as shown in Fig.\ref{Fig4}. The vertical $\rho=0$ side of the triangles then 
corresponds either to $r=\infty$ or to 
$r=-\infty$, which is the position of the timelike AdS boundary.

The null boundaries of the $A,B_\pm$ regions are the Killing horizons. 
They correspond to $\rho=\pm\infty$ but they can be reached by timelike
geodesics in a finite proper time. Specifically, the radial timelike geodesics 
are described by equation \eqref{geodesic} 
which can be represented in the equivalent form 
\be                            \label{gds}
\left(\frac{dr}{d\tau}\right)^2-\frac{{\cal E}^2}{\mu^2 Q^2} =-1,
\ee
where $\tau$ is the proper time. 
Let us denote $x=r-r_+$. 
The amplitude $Q$ has a simple zero at $r=r_+$ and close 
to this point one has $Q=\alpha x+{\cal O}(x^2)$
with a constant $\alpha$, 
in which case Eq.\eqref{gds}
yields  
\be                                       \label{xx}
x\,dx\propto d\tau~~~~~\Rightarrow~~~~~~~x^2\propto(\tau_0-\tau),
\ee
where $\tau_0$ is an integration constant. 
This shows that starting in the $A$ region where $x<0$, the 
geodesics arrive at the boundary where $x=0$ 
at a finite moment of the proper time, $\tau=\tau_0$.
Therefore, the $A$ region is geodesically incomplete and the geodesics 
arrive at its null boundary
in a finite proper time.  The same applies to the regions $B_\pm$. 

However, it turns out that the boundaries of the regions -- the Killing  horizons,
are singular since the curvature 
diverges  there. 
Introducing the orthonormal tetrad consisting 
of the vectors 
\be
e_0=\frac{1}{Q}\,\frac{\partial}{\partial t},~~~~~
e_1=\frac{\partial}{\partial r},~~~~~
e_2=\frac{1}{R}\,\frac{\partial}{\partial \vartheta},~~~~~
e_3=\frac{1}{R\sin\vartheta}\,\frac{\partial}{\partial \varphi},~~~~~
\ee
the following tetrad components of the curvature
do not vanish (with $^\prime=d/dr$) 
\bea                         \label{curv}
R_{0101}=\frac{Q^{\prime\prime}}{Q},~~
R_{0202}=R_{0303}
=\frac{Q^\prime R^\prime}{QR},~~R_{2323}=R_{2424}=-\frac{R^{\prime\prime}}{R},~~
R_{3434}=\frac{1-R^{\prime 2}}{R^2}.~~~~~
\eea
Since $Q\propto x$ at the horizon, the components 
$R_{0202}=R_{0303}\propto 1/x\propto 1/\sqrt{\tau_0-\tau}$ diverge, although  
this divergence is relatively mild and only leads to finite relative deviations
of the neighboring geodesics, and hence to finite tidal deformations.  
This can be seen by using the equation of the geodesic deviation and integrating
over the proper time $\tau$. However, the component 
$R_{0101}=Q^{\prime\prime}/Q$ also diverges
since 
$Q^{\prime\prime}\neq 0$ when $Q\to 0$, hence   
even the 2D metric $ds_g^2=-Q^2dt^2+d r^2$ is singular,
 even though  it reduces in the 
leading order to 
the flat Rindler metric $ds_g^2=-\alpha^2 x^2 dt^2+d r^2$.
Therefore, every radial geodesic approaching the horizon hits the curvature 
singularity.

One can nevertheless try and extend the geodesics beyond the horizon in a continuous way, 
which would allow one to construct a  Kruskal-type extension of the metric,
although only within the   ${\cal C}^0$ class. 
Geodesics of the extended metric, after having crossed the horizon, enter a 
``T-region'' where the $x^2$ in Eq.\eqref{xx} formally becomes negative, 
because space and time interchange their role and the metric becomes  
\bea                             \label{ansatz00000}
ds_g^2&=&+Q^2(r)dt^2-dr^2+R^2(r)d\Omega^2\,. 
\eea
Since $r$ becomes the timelike coordinate, this metric 
describes not the static wormhole  but
rather a dynamical cosmology, 
a minimum of $R(r)$ then  
corresponding not to the wormhole throat but rather to something like a cosmological 
bounce. A knowledge of such solutions would allow one  
to construct a maximal extension of the spacetime geometry in order to find out 
if the wormhole can be traversed by geodesics or not.  

However, solutions in the T-regions are presently not known. 
In addition, the maximal extension would only be continuous and not
differentiable due to the 
singular nature of the horizons (if there are solutions for which   
$Q^{\prime\prime}$ vanishes at the horizon
then their extension would be at least ${\cal C}^2$).   
We therefore do not pursue this line anymore and leave the problem
of constructing a maximal extension for 
the W1b wormhole geometry  for a future project.

\subsection{Type W2 wormholes}
Let us now consider solutions of the type shown in Fig.\ref{Fig2}.  
The g-geometry is globally regular and far away from the throat 
is described by Eq.\eqref{NQQQ}, 
so that in the leading order it approaches 
the AdS metric with the cosmological constant 
$\Lambda=-\kappa_1 b_0<0$. As a result, the structure of the 
g-geometry is essentially the same as for the type W1a solutions.  
The metric can be cast to the form \eqref{conf1}, where  
the conformal coordinate $\rho$ is defined by \eqref{rho}
and changes within a finite interval, $\rho\in(-\rho_\infty,+\rho_\infty)$.  
\begin{figure}[h]
\hbox to \linewidth{ \hss

	\psfrag{x}{$\rho=0$}
	\psfrag{xxx}{$\rho$}
		\psfrag{x1}{$\rho=\rho_\infty$}
	     \psfrag{x2}{$\rho=-\rho_\infty$}	

	\psfrag{light}{\large{light}}
	\psfrag{time}{\large{timelike geodesics}}
	\psfrag{f}{\Large{f-geometry}}
	\psfrag{g}{}

	\psfrag{t}{$t$}
	\psfrag{r}{$\rho$}
	\psfrag{I}{\Large{$J_{+}$}}
	\psfrag{II}{\Large{$J_{-}$}}

	\psfrag{Y}{$Y/Y_\infty$}
	\psfrag{N}{$N/N_\infty$}	

	\psfrag{Q}{$Q/Q_\infty$}
	\psfrag{q}{$\ln(q)$}
	\psfrag{U}{$\ln(U)$}	

	\psfrag{QQ}{$Q^2$}

	\resizebox{4cm}{6cm}
	{\includegraphics{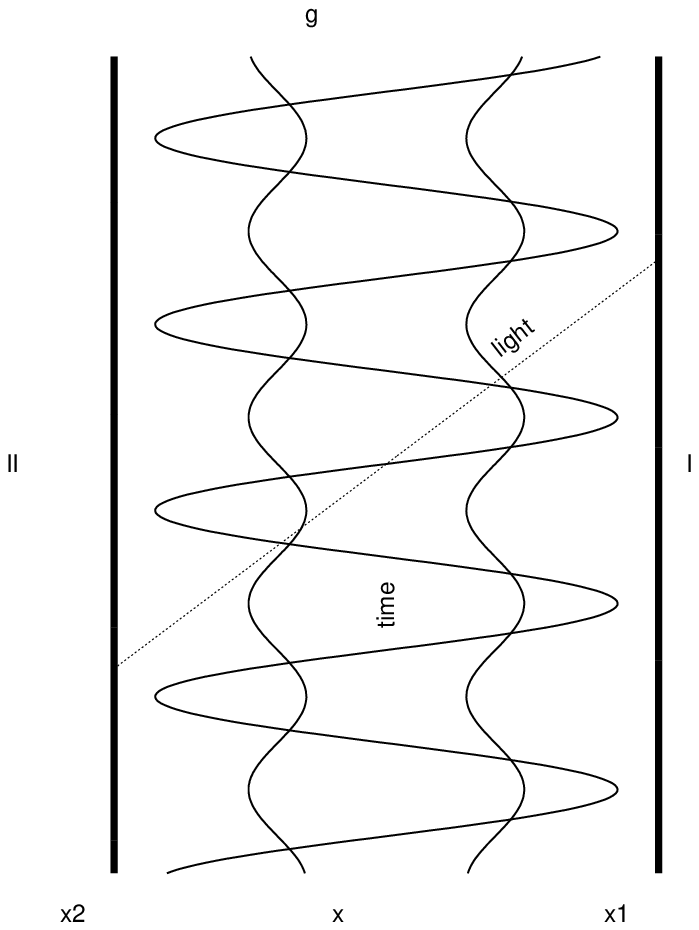}}
\hspace{20mm}
	\resizebox{6cm}{5cm}{\includegraphics{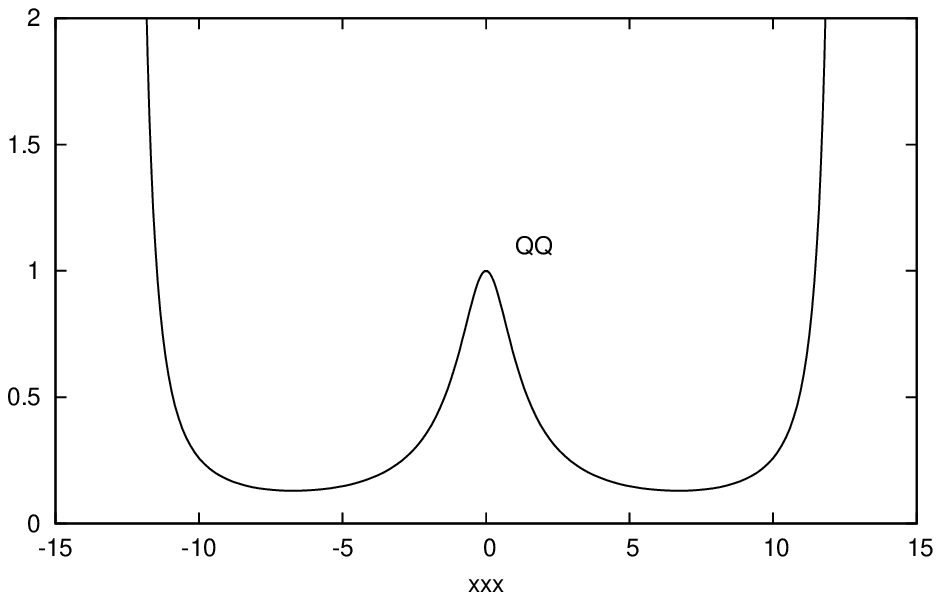}}
	
\hspace{1mm}
\hss}
\caption{{\protect\small 
The structure of the g-geometry 
for the W2 solution shown in Fig.\ref{Fig2} (left), and  
the corresponding effective potential $Q^2$   
in the geodesic equation \eqref{geodesic} (right).  
 }}%
 \label{Fig33}
\end{figure}
This gives the conformal diagram shown in Fig.\ref{Fig33}, which is similar to that
in Fig.\ref{Fig3}. The timelike geodesics are described by \eqref{geodesic}, whose 
potential $Q^2$ is shown in Fig.\ref{Fig33}. The wormhole throat 
is repulsive, and in addition there is an infinite repulsive barrier 
at the timelike boundary. As a result, 
particles with ${\cal E}<\mu$ oscillate between the throat and the boundary, 
while those with ${\cal E}>\mu$ traverse the throat and  oscillate between the
right and left boundaries, as shown  in Fig.\ref{Fig33}.

Let us now consider the f-geometry. 
Far away from the throat it is described by Eq.\eqref{NQQQ}, 
so that one has asymptotically 
\be                            \label{YYY1}
Y=const.\times R^2+{\cal O}(R),~~~~U=U_\infty+\frac{const.}{R}+{\cal O}\left(\frac{1}{R^2}\right),
\ee
with a constant $U_\infty$. 
Let us represent the metric as
\be 
ds_f^2=-q^2dt^2+\frac{dU^2}{Y^2}+U^2d\Omega^2=q^2(-dt^2+d\rho^2)+U^2d\Omega^2
\ee
with 
\be                 \label{xxx} 
\rho=\int_0^r \frac{U^\prime}{qY}\,dr. 
\ee
This radial coordinate changes within a finite range, $\rho\in (-\rho_\infty,\rho_\infty)$, 
because at large $r$ one has in view of \eqref{YYY1} 
$d\rho\propto {dR}/{R^4}$, 
therefore the integral in \eqref{xxx} converges at the upper limit
to a finite value $\rho_\infty$. The radial geodesics of the f-metric obey 
\be                            \label{gdsf}
\left(\frac{d\rho}{dt}\right)^2+\frac{\mu^2}{{\cal E}^2}\,q^2=1. 
\ee
As is seen in Fig.\ref{Fig2}, the $q$ amplitude interpolates between 
$q(0)$ and $q_\infty$. Therefore, for large enough ${\cal E}$ there 
are geodesics which cross the whole range of $\rho$ and arrive at $\rho=\pm \rho_\infty$ 
in a finite proper time. 
When $\rho\to \pm \rho_\infty$ 
the geometry becomes 
\be                              \label{fff}
ds_f^2=q_\infty^2(-dt^2+d\rho^2)+U_\infty^2 d\Omega^2\,,
\ee
which is completely regular. As a result, nothing prevents the f-geodesics from extending 
beyond the values $\rho=\pm \rho_\infty$. 
Hence, from the f-geometry viewpoint, the manifold 
corresponding to the interval $\rho\in (-\rho_\infty,\rho_\infty)$ is geodesically 
incomplete, so that the f-geometry could be extended beyond this interval. 
However, as far as the g-geometry is concerned, the manifold 
is complete, because the limiting values $\rho=\pm \rho_\infty$ correspond to the AdS 
boundary.    
We therefore have a peculiar situation where the same manifold is complete
in one geometry but is incomplete in the other. One could in principle try and extend the 
manifold by integrating the equations beyond  $\rho=\pm \rho_\infty$ until the f-geometry 
is complete. However, the additional parts of the manifold obtained in this way  
would then be g-geodesically disconnected from the original wormhole,
because the latter is already g-complete. 
We therefore adopt the viewpoint that only the g-metric 
describes the spacetime geometry, 
while the f-metric should be viewed as  a spin-2 tensor field 
whose geometric interpretation is possible but not necessary.

\section{Concluding remarks  \label{sec8} } 
\setcounter{equation}{0}

The above analysis gives strong (numerical) evidence in favor of the existence of wormholes 
in the bigravity theory.  These wormholes are very large, with the throat radius of the order of the 
inverse graviton mass, and they 
can be of two principal types, which we call W1 and W2. 

The W1 wormholes are asymptotically AdS. This feature can be understood by noting that 
the AdS space is an attractor at large $r$, which means the following. 
The solutions can be obtained by integrating the system of three first order 
equations \eqref{eqs} for $N(R),Y(R),U(R)$. 
At large $R$ the solutions approach the AdS values, so that 
\be                                        \label{linear} 
N=N_0\times(1+\nu),~~~~Y=N_0\times(1+\xi), ~~~~U=\lambda R\times(1+\chi), 
\ee
where $N_0^2=1-\Lambda R^2/3$ and the deviations $\nu,\xi,\chi$ are small. 
Let us consider first the W1a solution 
shown in Fig.\ref{Fig1}. Then one has $\Lambda=-0.170$ and 
$\lambda=0.358$ (see
Eqs.\eqref{lam1},\eqref{lam2}).
Linearizing the equations with respect to small $\nu,\xi,\chi$ then gives 
the solution
\be      \label{s}
\nu\sim\xi\sim \chi\sim R^s~~~~~\mbox{with}~~~s=-3,-\frac32\pm \omega\times i\,,
\ee
where, 
for the parameter values in \eqref{par}, one finds $\omega=2.068$.  
The three different values of $s$ correspond to three independent solutions,  
all of them 
approaching zero as $R\to\infty$. Therefore, the stable manifold around the AdS fixed point 
is three-dimensional, so that solutions of the three-dimensional system \eqref{eqs} generically
run into this fixed point, which is why this is an attractor. 
For comparison, the flat space is not an attractor since the stable manifold around it is only 
two-dimensional and  the solutions miss it, hence  they  
are not asymptotically flat 
(probably not even in exceptional cases; see below).

Eq.\eqref{s} determines the deviation from the AdS asymptotic, $\delta N^2=N^2-N_0^2$,
\be                            \label{sss}
\delta N^2=-\frac{2M}{R}+A\sqrt{R}\cos(\omega \ln(R)+\alpha), 
\ee
where $M,A,\alpha$ are integration constants. 
The first term on the right here is the  contribution of the massless 
graviton, while the second term is the effect of the scalar polarization of the massive graviton.
The embarrassing observation is that the massive contribution oscillates
(this is confirmed by the numerics) 
since $s$ given by \eqref{s} has a non-vanishing imaginary part $\omega$. This 
indicates that the mass is imaginary. Indeed, 
the Fierz-Pauli graviton mass 
for fluctuations around the proportional AdS background is given by 
(in units of $m$)  \cite{Hassan:2012wr}
\be                                 \label{BF}
m_{\rm FP}^2={\cal P}_1(\lambda)\left(\kappa_1\lambda+\frac{\kappa_2}{\lambda}\right).
\ee
For $\lambda=0.358$  this gives $m_{\rm FP}^2=-0.37$, 
hence the gravitons indeed behave as tachyons. 
The value of the graviton mass actually agrees with the value of $\omega$ given above, 
which can be seen by noting that the scalar graviton behaves as a scalar field. 
On the other hand,  
a static, spherically symmetric scalar field of mass $\mu$ on the AdS 
background decays asymptotically  as $R^s$ with  
\be                        \label{ss}
s=-\frac{3}{2}\pm \frac32\sqrt{1-\frac{4\mu^2}{3\Lambda}  }.
\ee
Setting here $\mu=0$ and choosing the minus sign yields $s=-3$, 
while setting $\mu^2=m_{\rm FP}^2=-0.37$ yields 
$s=-3/2\pm 2.068\times i$. This reproduces 
precisely the values in \eqref{s}.

At the same time, one should stress that  the very 
existence of tachyons in the AdS space is not necessarily a bad feature,
as long as their mass squared exceeds the BF bound 
$m^2_{\rm BF}=\frac{3}{4}\,\Lambda$ (the mass for which the square root
in \eqref{ss} vanishes), in which case they do not produce an instability 
\cite{Breitenlohner:1982bm}. 
However, for the W1a solution shown in Fig.\ref{Fig1} one has  
\be                        \label{BF1}
\mbox{type W1a:}~~~~~~~~m_{\rm FP}^2=-0.37<m^2_{\rm BF}\equiv \frac{3}{4}\,\Lambda=-0.12,
\ee 
therefore, the BF bound is violated, which implies 
that the solution is unstable.
It turns out that the BF bound is violated for all W1a solutions that we could find.  

On the other hand, for the W1b solution shown in Fig.\ref{Fig1a} one obtains 
\be                        \label{BF1a}
\mbox{type W1b:}~~~~~~~~m_{\rm FP}^2=-6.01>m^2_{\rm BF}\equiv \frac{3}{4}\,\Lambda=-6.36,
\ee 
so that the BF bound is fulfilled, therefore  the tachyon instability should be absent. 
This does not immediately imply that the W1b solutions are 
stable. However, since they do not suffer from the most dangerous instability, 
there is a chance that they could be stable, 
which however can only be decided  
after a special analysis.

Let us finally consider the W2 wormholes. They do not approach the proportional 
background and so it is less clear \cite{Hassan:2012wr} how to compute the Fierz-Pauli mass. 
However, the linearization of the field  equations around the asymptotic values, 
similar to that described by Eq.\eqref{linear}, gives for the deviations $\nu,\xi,\chi$
power law solutions with real powers. Therefore, there is no evidence for tachyons,
so that the W2 solutions could perhaps be stable. It should however be again 
emphasized that in all cases a detailed  stability analysis remains an open issue.

The tachyons \cite{Volkov:2014qca,Volkov:2014ida} and superluminal 
waves \cite{Deser:2012qx,Deser:2013eua,Deser:2014hga}
were previously detected in the massive gravity theory with a fixed f-metric. 
Their existence  does not necessarily mean that the theory is ill-defined but rather 
shows that it can have unphysical solutions.
It seems that in the bigravity theory the situation is similar -- solutions can be 
physical and unphysical \cite{Hassan:2014vja}. 
The described above W1a wormholes apparently belong to the latter 
category because they show tachyons and are unstable. 
One should also say that the solutions may admit a holographic interpretation, 
similarly to the massive gravity solutions used in the 
holographic conductivity models 
\cite{Blake:2013bqa,Amoretti:2014mma}.

It is instructive to compare the wormholes and black holes \cite{Volkov:2012wp}. In both cases 
one can use the Schwarzschild coordinate, 
$
ds^2_g=-Q^2dt^2+{dR^2}/{N^2}+R^2d\Omega^2\,.
$
For black holes both $N^2$ and $Q^2$ vanish at $R=h$ (horizon), while for wormholes 
$N^2$ vanishes at $R=h$ (throat) but $Q^2$ does not. The bigravity black holes  
\cite{Volkov:2012wp} are characterized 
by two independent values,
$h$ and $\sigma=U(h)/h$, and 
they can be obtained by integrating 
Eqs.\eqref{eqs} for $N(R),Y(R),U(R)$ with the boundary condition $N(h)=Y(h)=0$.
The equation 
 $Q^\prime=FQ$ (Eq.\eqref{Q1}) then insures that $Q(h)=0$, since one generically has 
at the horizon 
$2F=1/(R-h)+{\cal O}(1)$.  
Now, the wormholes are actually the same solutions but obtained for special values 
of $\sigma$ (given by Eq.\eqref{D1}) 
for which the pole of $F$ is canceled, and so the equation $Q^\prime=FQ$  ensures that 
 $Q$ is finite as $R=h$.
From this viewpoint, wormholes can be viewed as the special case of 
black holes corresponding to the fine-tuned $\sigma$. 

The bigravity black holes generically approach the AdS space \cite{Volkov:2012wp},
but in exceptional cases, for specially adjusted values 
of $\sigma$ (and for $h>0.86$),  they can be asymptotically flat 
\cite{Brito:2013xaa}.
For the wormholes the value of $\sigma$ is already fixed by the 
condition of having a regular throat, so that one cannot further adjust  it 
to fulfill the asymptotic flatness condition as well. Therefore, 
asymptotically flat wormholes are unlikely to exist.

The symmetric wormholes exist only in the bigravity 
and not in the massive gravity theory with a flat f-metric. Indeed, the 
flat f-metric requires that $Y=1$, which is  not compatible 
with the boundary condition expressed by  \eqref{YYY}. However, we have checked 
that in the massive gravity limit there are non-symmetric under $r\to -r$ 
wormhole-type solutions
for which $R$ develops a minimum, and even infinitely many minima.

\acknowledgments
We are grateful to Eugen Radu and Jeorge Rocha and especially to Gary Gibbons
for discussions and constructive suggestions.  
This work was partly supported by the Russian Government Program of Competitive Growth 
of the Kazan Federal University and also by Grant 14-02-00598
of the Russian Foundation for Basic Research. 



\providecommand{\href}[2]{#2}\begingroup\raggedright\endgroup

\end{document}